\documentclass[twocolumn,letterpaper,aps,prl,superscriptaddress,showpacs,amsmath]{revtex4-1}

\usepackage{graphicx}

\usepackage{txfonts}\usepackage[full]{textcomp}
\newcommand{\vc}[1]{\boldsymbol{#1}}
\newcommand{\pd}{{\phantom{\dagger}}}

\begin{document}

\title{\fontsize{12.2}{14}\selectfont Raman scattering from Higgs mode oscillations in the two-dimensional antiferromagnet Ca$_\mathbf{2}$RuO$_{\mathbf 4}$} 

\author{Sofia-Michaela Souliou}
\affiliation{Max Planck Institute for Solid State Research,
Heisenbergstrasse 1, D-70569 Stuttgart, Germany}
\affiliation{European Synchrotron Radiation Facility, 
BP 220, F-38043 Grenoble Cedex, France}

\author{Ji\v{r}\'{\i} Chaloupka}
\affiliation{Central European Institute of Technology,
Masaryk University, Kamenice 753/5, 62500 Brno, Czech Republic}
\affiliation{Department of Condensed Matter Physics, Faculty of Science,
Masaryk University, Kotl\'a\v{r}sk\'a 2, 61137 Brno, Czech Republic }

\author{Giniyat Khaliullin}
\affiliation{Max Planck Institute for Solid State Research,
Heisenbergstrasse 1, D-70569 Stuttgart, Germany}

\author{Gihun~Ryu}
\affiliation{Max Planck Institute for Solid State Research,
Heisenbergstrasse 1, D-70569 Stuttgart, Germany}

\author{Anil~Jain}
\affiliation{Max Planck Institute for Solid State Research,
Heisenbergstrasse 1, D-70569 Stuttgart, Germany}

\author{B.~J.~Kim}
\affiliation{Max Planck Institute for Solid State Research,
Heisenbergstrasse 1, D-70569 Stuttgart, Germany}

\author{Matthieu Le Tacon}
\affiliation{Max Planck Institute for Solid State Research,
Heisenbergstrasse 1, D-70569 Stuttgart, Germany}
\affiliation{Karlsruhe Institute of Technology, 
Institut f\"{u}r Festk\"{o}rperphysik, D-76021 Karlsruhe, Germany}

\author{Bernhard Keimer}
\affiliation{Max Planck Institute for Solid State Research,
Heisenbergstrasse 1, D-70569 Stuttgart, Germany}

\begin{abstract}\fontsize{9.5}{11.2}\selectfont 

We present and analyze Raman spectra of the Mott insulator Ca$_2$RuO$_4$,
whose quasi-two-dimensional antiferromagnetic order has been described as a
condensate of low-lying spin-orbit excitons with angular momentum
$J_\mathrm{eff}=1$. In the $A_g$ polarization geometry, the amplitude (Higgs)
mode of the spin-orbit condensate is directly probed in the scalar channel,
thus avoiding infrared-singular magnon contributions. In the $B_{1g}$
geometry, we observe a single-magnon peak as well as two-magnon and two-Higgs
excitations. Model calculations using exact diagonalization quantitatively
agree with the observations. Together with recent neutron scattering data, our
study provides strong evidence for excitonic magnetism in Ca$_2$RuO$_4$ and
points out new perspectives for research on the Higgs mode in two dimensions.

\end{abstract}

\date{\today}


% 75 Magnetic properties and materials
% 75.10.Jm Quantized spin models, including quantum spin frustration

~\maketitle\fontsize{10.75}{12.4}\selectfont

The notion of Goldstone and Higgs modes, corresponding to phase and amplitude
oscillations of a condensate of quantum particles, appears in many areas of
physics including magnetism \cite{Pek15}. In quantum magnets, especially near
quantum criticality \cite{Sac11}, the magnetization density is far from being
saturated and hence allowed to oscillate near its mean value, forming a
collective amplitude mode.

The ``magnetic'' Higgs mode has been observed \cite{Rue08} in quantum dimer
systems, where the magnetic order is due to Bose-Einstein condensation of
spin-triplet excitations \cite{Gia08}. A conceptually similar, but physically
distinct case is expected in Van Vleck-type Mott insulators, where the
``soft'' moments result from condensation of spin-orbit excitons \cite{Kha13},
that is, magnetic transitions between spin-orbit $J_\mathrm{eff}=0$ and
$J_\mathrm{eff}=1$ levels propagating via exchange interactions. Recent
inelastic neutron scattering (INS) experiments \cite{Jai17} on Ca$_2$RuO$_4$
have indeed revealed Higgs oscillations of the magnetization in this material,
which is based on nominally non-magnetic, spin-orbit singlet Ru$^{4+}$ ions. 
A detailed analysis of the dispersion relations of the Higgs mode and magnons
determined by INS showed that Ca$_2$RuO$_4$ is close to a quantum critical
point associated with the condensation of $J_\mathrm{eff}=1$ excitons
\cite{Jai17}.

The unique aspect of Ca$_2$RuO$_4$ is that it hosts Higgs physics in a
two-dimensional setting, which has been a focus of many theoretical studies
\cite{Pod11,Pod12,Pol12,Che13,Gaz13a,Gaz13b,Ran14,Ros15,Kat15}. As the
magnetization density is not a conserved quantity, the Higgs mode is not
symmetry protected, and various decay processes convert it into a many-body
resonance with $\sim\omega^3$ onset. It was also emphasized \cite{Pod11} that
the actual appearance of this resonance strongly depends on the symmetry of
the probe. In INS experiments, which probe the longitudinal magnetic
susceptibility, the low-energy behaviour of the Higgs resonance is masked by
the infrared-singular two-magnon contribution. To avoid contamination by the
Goldstone modes, the probe should couple to the condensate in the scalar
channel (i.e. insensitively to the phase/direction). Precisely this type of
experiment has been done in ultracold atomic systems \cite{End12}. 

In this Letter, we demonstrate that Raman light scattering in the fully
symmetric, i.e. $A_g$ channel can serve as a scalar probe in magnetic systems,
thus providing direct access to Higgs oscillations of ``soft'' moments. While
in conventional Heisenberg magnets with rigid spins (such as La$_2$CuO$_4$ or
Sr$_2$IrO$_4$) the $A_g$ channel is magnetically silent, the size of the local
moments, and hence the magnetization density in excitonic systems is
determined by a balance between the spin-orbit $\lambda$ and exchange
$J$-interactions \cite{Kha13,Jai17}, and the $A_g$ modulation of the latter
directly shakes the condensate density.

The Raman scattering data in Ca$_2$RuO$_4$ presented below indeed reveal a
pronounced magnetic contribution in the $A_g$ channel, which we identify and
describe using the same excitonic model that has already been parameterized in
the INS study \cite{Jai17}. In the $B_{1g}$ channel, we observe the expected
two-magnon scattering and an additional two-Higgs scattering contribution, as
well as a single-magnon peak. All the observations are coherently explained by
model calculations.

%======================================================================

\begin{figure}[tb] 
\includegraphics[width=8.4cm]{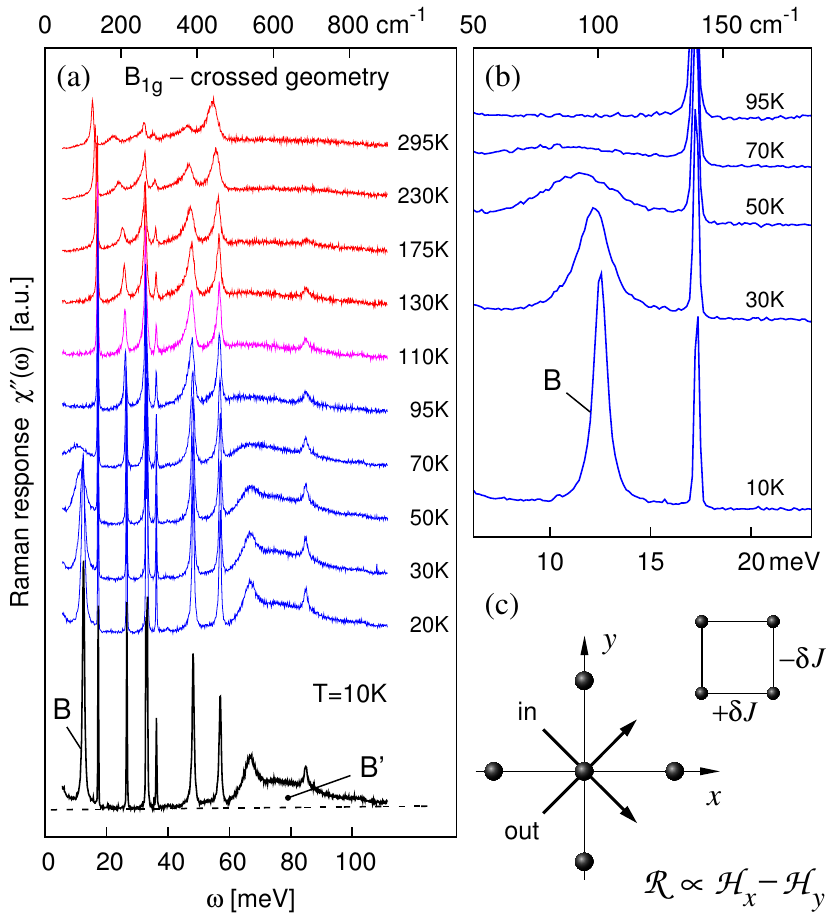} 
\caption{\fontsize{9.8}{11.2}\selectfont
(a)~Raman spectra in $B_{1g}$ scattering geometry with two magnetic features
$B$, $B'$ appearing below $T_N$. The background (dashed line) is subtracted in
further analysis.
(b)~Detailed view on the feature $B$.
(c)~Polarization vectors of incoming and outgoing photons with respect to the
Ru lattice. In the Raman process, the exchange is modulated with opposite signs
on $x$ and $y$ bonds leading to the Raman operator 
$\mathcal{R}\propto\mathcal{H}_x-\mathcal{H}_y$.
}
\label{fig:experBg}
\end{figure}

\begin{figure}[tb] 
\includegraphics[width=8.4cm]{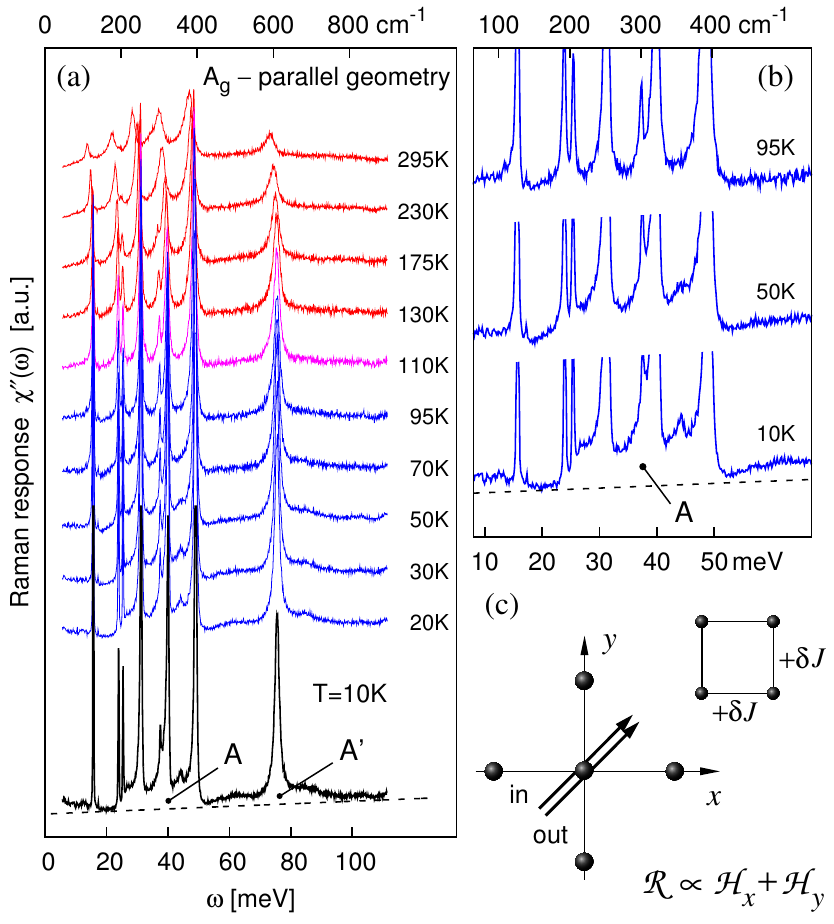} 
\caption{\fontsize{9.8}{11.2}\selectfont
(a)~Raman spectra in $A_g$ scattering geometry with two magnetic 
features $A$, $A'$ appearing below $T_N$.
(b)~Detailed view on the feature $A$.
(c)~The polarization vectors in the $A_g$ setup. In this case, the 
exchange is modulated equally on $x$ and $y$ bonds and the Raman operator 
$\mathcal{R}\propto\mathcal{H}_x+\mathcal{H}_y$. 
}
\label{fig:experAg}
\end{figure}

{\it Experiment.---}Single crystals of Ca$_2$RuO$_4$ with
$T_N=110\:\mathrm{K}$ were grown by a floating zone method, as described
elsewhere \cite{Nak01}. The Raman data were recorded on a Labram (Horiba
Jobin-Yvon) single-grating Raman spectrometer, using the
$632.817\:\mathrm{nm}$ line of a He$^+$/Ne$^+$ mixed gas laser. The
experiments were performed in backscattering geometry along the
crystallographic \mbox{$c$-axis}. Ca$_2$RuO$_4$ crystalizes in the
orthorhombic \mbox{\textit{Pbca}-\textit{D}$_{2\textit{h}}^{15}$} space group.
Excitations in the $B_{1g}$ and $A_g$ representations of the point group
\textit{D}$_{2\textit{h}}$ were probed in crossed and parallel configurations
respectively, with the polarization of the incident light at $45^\circ$ to the
Ru-Ru bonds [see Figs.~\ref{fig:experBg}(c) and \ref{fig:experAg}(c)].  The
spectra were corrected for the Bose thermal factor to obtain the Raman
response functions $\chi''(\omega)$.

% Fig.1 -------------------------------------------------------------------
%--------------------------------------------------------------------------

% Fig.2 -------------------------------------------------------------------
%--------------------------------------------------------------------------

Temperature-dependent $\chi''(\omega)$ in the range of $5$ to
$110\:\mathrm{meV}$ are plotted in Figs.~\ref{fig:experBg} and
\ref{fig:experAg}. The frequencies of the observed phonon modes are in good
agreement with previous Raman studies \cite{Rho05}. The phonon modes are
superimposed on top of a broad continuum. As the temperature is
lowered, the continuum evolves into distinct spectral features $B$, $B'$
(Fig.~\ref{fig:experBg}) and $A$, $A'$ (Fig.~\ref{fig:experAg}). The
temperature dependence of the new features follows closely that of the
magnetic order parameter and strongly suggests their magnetic origin. The
fact that these excitations are well inside the optical gap exceeding
$0.5\:\mathrm{eV}$~\cite{Jun03} further supports this interpretation.

More specifically, in the $B_{1g}$ channel, the feature $B$ appears around
$12\:\mathrm{meV}$ and gradually sharpens [Fig.~\ref{fig:experBg}(b)].
Earlier Raman studies attributed it either to two-magnon
scattering~\cite{Sno02,Rho03} or to a zone-boundary folded phonon in the
magnetically ordered state~\cite{Rho05}. 
However, we find below that the two-magnon scattering is represented by the
$B'$ structure around $80\:\mathrm{meV}$, while the $B$-peak is 
identified as a single-magnon excitation.

In the $A_{g}$ channel, the $A$-structure in the range of $25$ to
$50\:\mathrm{meV}$ develops in the magnetically ordered state
[Fig.~\ref{fig:experAg}(b)]. The phonon modes in this spectral region exhibit
pronounced Fano-type asymmetric lineshapes -- a~clear signature of the
presence of a continuum of excitations coupled to the phonons. As noticed
above, the large optical gap implies a magnetic origin of the continuum.

%======================================================================

{\it Extraction of the magnetic response.---}We adopt the Green's function 
approach \cite{Nit74,Kle75,Che93} to the Raman
response of the coupled system of phonons and a continuum. We describe the
system by a matrix propagator whose inverse $G^{-1}(\omega)$ contains the
response functions of the magnetic
$[G^{-1}(\omega)]_{00}=R(\omega)+iS(\omega)$
and phonon $[G^{-1}(\omega)]_{nn}=\omega_n-\omega-i\Gamma_n$
($n=1\ldots N$) subsystems as the diagonal elements. The coupling between
phonon $n$ and the continuum is provided by nondiagonal matrix
elements $[G^{-1}(\omega)]_{n0}=[G^{-1}(\omega)]_{0n}=V_n$. After inverting
$G^{-1}(\omega)$, the Raman response is obtained as
$\chi''(\omega)=\sum_{j=0}^N W_j[\mathrm{Im}\,G(\omega)]_{jj}$, where 
$W_j$ are spectral weights of the normal modes of the coupled spin-phonon 
system.

% Fig.3 -------------------------------------------------------------------
\begin{figure}[tb] 
\includegraphics[width=8.4cm]{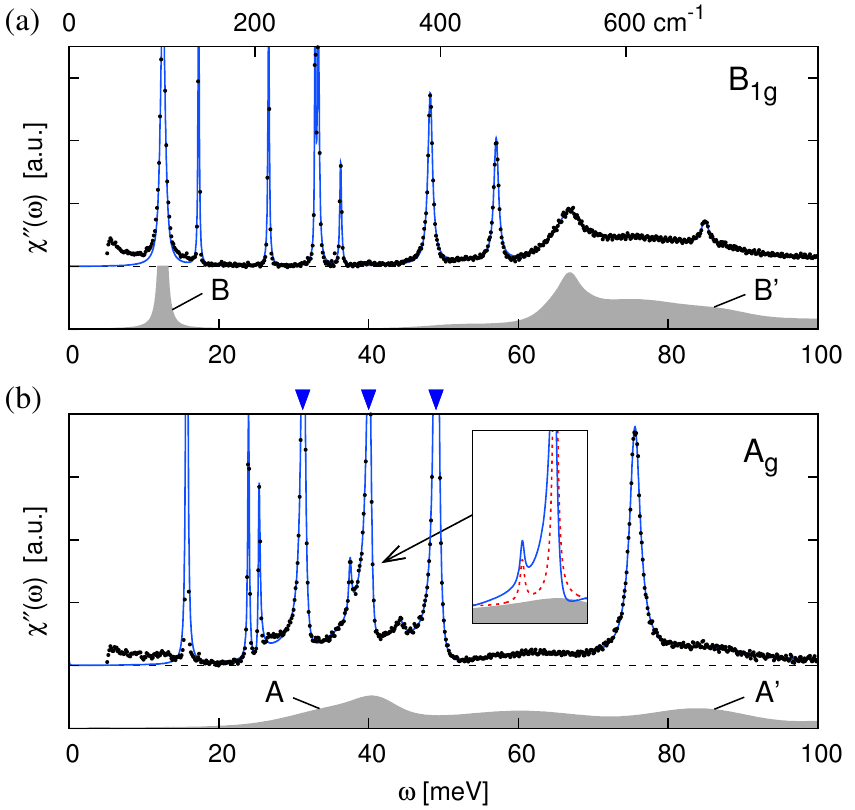} 
\caption{\fontsize{9.8}{11.2}\selectfont
Fits of $T=10\:\mathrm{K}$ Raman spectra in $B_{1g}$ (a) and $A_g$ (b)
channels using a model of phonons interacting with a magnetic continuum. The
model response (blue) is compared to the experimental points (black). The
obtained magnetic signal $S(\omega)$ is indicated by shading. The $A_g$
phonons marked by blue triangles are most strongly affected by the spin-phonon
interaction which changes their lineshape dramatically, compared to the
non-interacting case (red dashed line in the inset). The associated
spectral-weight transfer is moderate only.
}
\label{fig:fit}
\end{figure}
%--------------------------------------------------------------------------

The magnetic response functions $S(\omega)$, determined by
fitting $\chi''(\omega)$ to the low-temperature spectra, are presented in
Fig.~\ref{fig:fit}. While in the $B_{1g}$ case the above procedure just
confirms the expected result, in the $A_g$ case it proved essential to obtain
the actual $S(\omega)$ profile. The feature $A$ is found to be peaked at about
$40\:\mathrm{meV}$ and has a long tail that merges with the high-energy
continuum ($A'$), much flatter than the $B_{1g}$ one ($B'$).

%======================================================================

% Fig.4 -------------------------------------------------------------------
\begin{figure}[tb] 
\includegraphics[width=8.2cm]{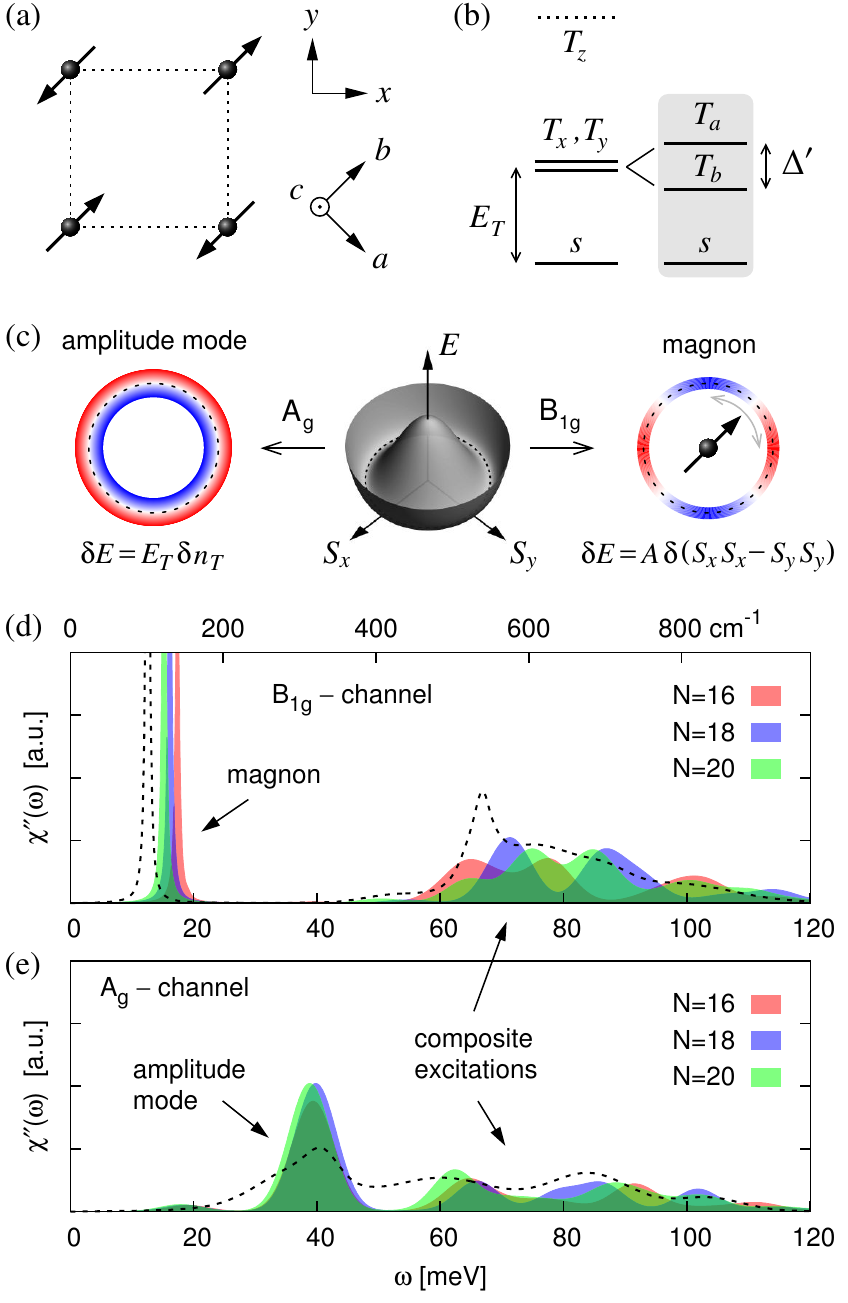} 
\caption{\fontsize{9.8}{11.2}\selectfont
(a)~Coordinate frames for the Ru lattice. The ordered moments point 
along the $b$ axis.
(b)~Multiplet structure of Ru$^{4+}$ ions with groundstate singlet $s$ with
$J_\mathrm{eff}=0$ and higher-lying magnetic states derived from
$J_\mathrm{eff}=1$ triplet. A~tetragonal crystal field removes the 
degeneracy of the triplet states $T$ by lifting up $T_z$ \cite{Jai17}. An 
orthorhombic distortion further splits $T_{x/y}$ into 
$T_{a/b}=\frac1{\sqrt2}(T_x\mp T_y)$ levels
forming together with $s$ the basis for the low-energy model (shaded).
(c)~Modulation of the condensate energy (Mexican-hat potential) in the Raman
process leading to an excitation of the amplitude mode ($A_g$ channel). Note
that $\delta n_T\equiv\delta (S_x^2+S_y^2)$, i.e. the $A_g$-coupling is
rotationally invariant. In contrast, the $B_{1g}$-coupling leads to the shape
deformations, leaving condensate density intact. A single magnon is excited
instead of the amplitude mode.
(d),(e)~Raman spectra obtained by exact diagonalization on clusters with
$N=16$, $18$, and $20$ sites \cite{noteBetts} using $E_T=31\:\mathrm{meV}$,
$J=7.5\:\mathrm{meV}$, $A=2.3\:\mathrm{meV}$, $\alpha=0.15$, and 
$\Delta'=4\:\mathrm{meV}$. The ED data for $B_{1g}$ (a) and $A_g$ (b) 
channels are presented in identical scales and overlayed by the magnetic 
$S(\omega)$ from Fig.~\ref{fig:fit}(a),(b) (dashed).
}
\label{fig:model}
\end{figure}
%--------------------------------------------------------------------------

{\it Magnetic model.---}In the following, we give a quantitative
interpretation of the magnetic features using the excitonic model of
Ref.~\cite{Kha13}, refined further by a comparison to INS data \cite{Jai17}.
The model utilizes the local basis depicted in Fig.~\ref{fig:model}(b)
stabilized by intraionic spin-orbit coupling. 
The dominant energy scale corresponds to the energy cost $E_T$ of a triplon
$T$ (derived from $J_\mathrm{eff}=1$ states) relative to that of the singlet
ground state $s$ ($J_\mathrm{eff}=0$). Its competition with the spin-orbital
exchange interaction results in a quantum critical point separating the
paramagnetic phase (dilute ``gas'' of $T$ on top of $s$ background) and
antiferromagnetic phase (condensate with coherently mixed $T$ and $s$). In
terms of hardcore bosons $s$ and $T_{x/y}$ associated with the relevant
low-energy levels and obeying local constraint $n_s+n_T=1$, these main
constituents of the model are expressed as
\begin{equation}\label{eq:HETJ}
\mathcal{H} = E_T\! \sum_i n_{T_i} +
J\!\!\!\!\! \sum_{\langle ij\rangle, \gamma=x,y} \!\!\!
\left( T^\dagger_{\gamma i} s^{\pd}_i s^\dagger_j T^{\pd}_{\gamma j}
- T^\dagger_{\gamma i} s^{\pd}_i T^\dagger_{\gamma j} s^{\pd}_j 
+ \mathrm{H.c.} \right).
\end{equation}
The exchange interaction $J$ comprises triplon hopping and pair
creation/annihilation which act together to form AF-aligned pairs
of Van Vleck moments.

The full model is most conveniently expressed using pseudospin $S$=1 formed by
the three levels $\{s,T_x,T_y\}$ \cite{Jai17}. The corresponding in-plane
operators
$S_\gamma=-i(s^\dagger T^{\pd}_\gamma-T^\dagger_\gamma s)$ for $\gamma=x,y$ 
are directly linked to the dominating Van Vleck part of magnetic moment,
while $S_z=-i(T^\dagger_x T^{\pd}_y-T^\dagger_y T^{\pd}_x)$ is related to
the moment residing in the excited $T$ levels. In this basis, the $J$-term in 
Eq.~\eqref{eq:HETJ} takes a form of the XY-model $J(S^x_i S^x_j+S^y_i S^y_j)$.
Supplemented by the bond-directional interaction $A$ and coupling between the
out-of-plane $S_z$ components, the exchange Hamiltonian for the $x$-bonds 
reads as
\begin{equation}
\mathcal{H}_x =\! \sum_{\langle ij\rangle\,\parallel\,x}
\left[ (J\!+\!A)\, S_i^x S_j^x + (J\!-\!A)\, S_i^y S_j^y +
J(1\!-\!\alpha)\, S_i^z S_j^z \right].
\end{equation}
The signs of the $A$ terms are opposite for $y$ bonds. The $T_{x/y}$-level
orthorhombic splitting [see Fig.~\ref{fig:model}(a),(b)] orienting the moments 
along $b$ axis translates into a single-ion anisotropy 
$\mathcal{H}_{\Delta'}=-\Delta'(S_x S_y+S_y S_x)=\frac12\Delta'(S_a^2-S_b^2)$. 
The full Hamiltonian used below is then
$\mathcal{H}=E_T n_{T}+\mathcal{H}_x+\mathcal{H}_y+\mathcal{H}_{\Delta'}$,
with $n_T=S_z^2$. 

%======================================================================

{\it Model calculations and interpretation of the data.---}We employ the
Loudon-Fleury \cite{Fle68} Raman scattering operator
$\mathcal{R}\propto\sum_{\langle ij\rangle}
(\vc{\epsilon}_\mathrm{in}\cdot\vc{r}_{ij})
(\vc{\epsilon}_\mathrm{out}\cdot\vc{r}_{ij})\,\mathcal{H}_{ij}$, which
modulates the exchange interactions $\mathcal{H}_{ij}$ in a way determined by
the incoming $\vc{\epsilon}_\mathrm{in}$ and outgoing
$\vc{\epsilon}_\mathrm{out}$ polarization vectors \cite{notezz}.
Specifying $\vc{\epsilon}_\mathrm{in}$ ($\vc{\epsilon}_\mathrm{out}$) 
by its angle $\varphi$ ($\varphi'$) to the $a$ axis, $\mathcal{R}$ becomes
\begin{equation}\label{eq:R}
\mathcal{R} \propto 
 \cos(\varphi-\varphi')\, (\mathcal{H}_x+\mathcal{H}_y)
+\sin(\varphi+\varphi')\, (\mathcal{H}_x-\mathcal{H}_y) .
\end{equation}
For $B_{1g}$ ($\varphi=0$, $\varphi'=\pi/2$) and $A_g$ 
($\varphi=\varphi'=\pi/2$) symmetries, only the $\mathcal{H}_x-\mathcal{H}_y$
or $\mathcal{H}_x+\mathcal{H}_y$ term above is active, respectively.

We first discuss the implications of Eq.~\eqref{eq:R} on a qualitative level.
Consider the $A_{g}$ scattering channel with
$\mathcal{R}\propto\mathcal{H}_x+\mathcal{H}_y$. While in the usual rigid spin
systems (e.g. cuprates) this operator is proportional to the Hamiltonian
itself and does not bring any dynamics, here we may replace it by its complement 
in the Hamiltonian, i.e. $\mathcal{R}\propto E_T n_T$ (and a small $\Delta'$ 
term), and obtain a non-trivial spectrum. Most importantly, $E_T n_T$ globally
changes the balance between the $s$ and $T_{x/y}$ components coherently mixed
in the condensate, exciting thus directly the amplitude mode of the
condensate. This $A_g$ Raman process may be intuitively understood as a
forced expansion and contraction of the Mexican-hat potential in
Fig.~\ref{fig:model}(c). In contrast to INS, the amplitude mode is probed
here in a rotationally invariant way, using a scalar coupling to the
condensate density. We thus avoid the contamination by the two-magnon
response that leads to a drastic broadening of the longitudinal mode in the
dynamical spin susceptibility.

In the $B_{1g}$ channel, the modulation of the exchange $J$ contained in
$\mathcal{R}\propto\mathcal{H}_x-\mathcal{H}_y$ produces a high-energy
two-magnon continuum, as in usual Heisenberg magnets. Here it is additionally
supported by other composite excitations such as two-Higgs continuum (similar
to what found in a soft-spin model \cite{Wei15}). A special role is played by
the bond-anisotropic $A$ term contributing to $\mathcal{R}$ as 
$A\sum_{\langle ij\rangle}(S^x_i S^x_j-S^y_i S^y_j)$. The resulting 
quadrupolar modulation of the condensate energy [see Fig.~\ref{fig:model}(c)]
drives the ordered moment toward the $x$ or $y$ directions hence exciting a
magnon.

To confirm the above expectations and make a quantitative comparison to the
experiment, in Fig.~\ref{fig:model}(d),(e) we show Raman spectra calculated by
exact diagonalization (ED). The best fit to the magnetic intensity extracted
in Fig.~\ref{fig:fit} is obtained for the parameters $E_T=31\:\mathrm{meV}$,
$J=7.5\:\mathrm{meV}$, $A=2.3\:\mathrm{meV}$, $\alpha=0.15$, and
$\Delta'=4\:\mathrm{meV}$, well matching those from the INS data~\cite{Jai17}.
The small differences in $E_T$ and $J$ is due to the different methods -- the
spin-wave approach~\cite{Jai17} versus ED used here.

In accord with the above discussion, the $B_{1g}$ model spectrum in
Fig.~\ref{fig:model}(d) contains a high-energy continuum and a single-magnon
peak due to the bond-directional $A$-part of $\mathcal{R}$ that
sums up to $A\sum_{\langle ij\rangle}(S^a_i S^b_j+S^a_i S^b_j)$.
Approximating $S$ along the ordered moment direction by 
$S^b_{\vc{R}}\approx\langle S_\parallel\rangle\,\mathrm{e}^{i\vc{Q}\cdot\vc{R}}$
with $\vc{Q}=(\pi,\pi)$, this part becomes 
$A\langle S_\parallel\rangle\,S^a_{\vc{Q}}$ 
thus probing the magnon at the ordering vector. The energy of the experimental
feature $B$ of about $12.5\:\mathrm{meV}$ indeed agrees with that of INS
$(\pi,\pi)$-magnon peak \cite{Jai17,Kun15}. The spectral weight (SW) of the
peak $B$ is roughly proportional to $A^2$, enabling us to estimate $A$ by
comparing the SW of $B$ and that of the $B'$ continuum. The
experimental SW ratio obtained from Fig.~\ref{fig:fit}(a) amounts to $0.27$.
In the model calculations, the average through the three clusters gives a
consistent value of $0.30$, confirming $A\simeq 2.3\:\mathrm{meV}$ taken from
INS fits. 

In the $A_g$ channel, the model spectrum in Fig.~\ref{fig:model}(e) is
dominated by the amplitude mode appearing at $40\:\mathrm{meV}$ in agreement
with the expected position of the bare amplitude mode based on INS (see Fig.~4
of Ref.~\cite{Jai17}). The amplitude mode peak is accompanied by a high-energy
continuum [Fig.~\ref{fig:model}(e)]. Since it is a part of the $n_T$
susceptibility, its profile is rather different than that of the (mainly)
two-magnon continuum in the $B_{1g}$ channel. The limited scattering
possibilities on the small clusters do not allow us to access the mode profile
by ED in detail. The available results for the relativistic quantum $O(N)$
model in $2+1$ dimensions \cite{Gaz13a,Gaz13b,Ran14,Ros15} suggest a Higgs
peak with $\sim\omega^3$ onset and an extended tail which is in a qualitative
agreement with $S(\omega)$ extracted in Fig.~\ref{fig:fit}(b).

Finally, we comment on the notable interplay of phonons with the amplitude
mode observed in Fig.~\ref{fig:fit}(b). First, $A_g$ phonons involving
rotations and tiltings of RuO$_6$ octahedra modify the Ru-O-Ru bond angle,
thus modulating the exchange $J$ in a symmetric fashion. Second, deformations
of the octahedra affect the splitting among $t_{2g}$ orbitals, thus modulating
$E_T$ owing to the different orbital composition of $s$ and $T_{x/y}$ states.
Both mechanisms provide a natural coupling of phonons to oscillations of the
condensate density that is determined by the ratio $E_T/J$.

%======================================================================

In conclusion, we have presented Raman light scattering data on Ca$_2$RuO$_4$
and fully interpreted its magnetic features in terms of the excitonic model
\cite{Kha13,Jai17}. As demonstrated, the $A_g$ scattering channel enables
direct access to the amplitude (Higgs) mode of the spin-orbit condensate. In
contrast to INS, the Higgs mode is probed here via a scalar coupling and is
not obscured by the two-magnon continuum. The overall agreement with both the
neutron and Raman experiments strongly supports the excitonic picture as the
basis for magnetism of Ca$_2$RuO$_4$. More generally, our results encourage
future experimental efforts to explore other compounds based on Van Vleck-type
ions such as Ru$^{4+}$, Os$^{4+}$, and Ir$^{5+}$.

JC acknowledges support by the Czech Science Foundation (GA\v{C}R) under Project
No. GJ15-14523Y and M\v{S}MT \v{C}R under NPU II project CEITEC 2020 (LQ1601).
BK acknowledges support by the European Research Council under Advanced Grant
669550 (Com4Com) and by the German Science Foundation (DFG) under the
coordinated research program SFB-TRR80.


\begin{thebibliography}{99}\fontsize{9.6}{11.0}\selectfont

% Higgs in condensed matter review
\bibitem{Pek15}
D.~Pekker and C.~M.~Varma, 
Annu. Rev. Condens. Matter Phys.~{\bf 6}, 269 (2015).

\bibitem{Sac11}
S.~Sachdev and B.~Keimer, Phys. Today {\bf 64}, 29 (2011).

% Higgs in dimer system
\bibitem{Rue08}
Ch.~R\"{u}egg, B.~Normand, M.~Matsumoto, A.~Furrer, D.~F.~McMorrow, 
K.~W.~Kr\"{a}mer, H.-U.~G\"{u}del, S.~N.~Gvasaliya, H.~Mutka, 
and M.~Boehm, Phys. Rev. Lett. {\bf 100}, 205701 (2008).

\bibitem{Gia08}
T.~Giamarchi, Ch.~R\"{u}egg, and O.~Tchernyshyov, 
Nature Phys. {\bf 4}, 198 (2008).

% original model
\bibitem{Kha13}
G.~Khaliullin, Phys. Rev. Lett. {\bf 111}, 197201 (2013).

% our INS
\bibitem{Jai17}
A.~Jain, M.~Krautloher, J.~Porras, G.~H.~Ryu, D.~P.~Chen, D.~L.~Abernathy, 
J.~T.~Park, A.~Ivanov, J.~Chaloupka, G.~Khaliullin, B.~Keimer, and B.~J.~Kim,
Nature Phys. {\bf 13}, 633 (2017).

% scalar probe vs longitudinal susceptibility
\bibitem{Pod11}
D.~Podolsky, A.~Auerbach, and D.~P.~Arovas,
Phys. Rev. B {\bf 84}, 174522 (2011).

% Higgs spectral function from large N expansion
\bibitem{Pod12}
D.~Podolsky and S.~Sachdev,
Phys. Rev. B {\bf 86}, 054508 (2012).

% Bose-Hubbard, QMC
\bibitem{Pol12}
L.~Pollet and N.~Prokof'ev,
Phys. Rev. Lett. {\bf 109}, 010401 (2012).

\bibitem{Che13}
K.~Chen, L.~Liu, Y.~Deng, L.~Pollet, and N.~Prokof'ev,
Phys. Rev. Lett. {\bf 110}, 170403 (2013).

% Monte Carlo near QCP
\bibitem{Gaz13a}
S.~Gazit, D.~Podolsky, and A.~Auerbach,
Phys. Rev. Lett. {\bf 110}, 140401 (2013).

\bibitem{Gaz13b}
S.~Gazit, D.~Podolsky, A.~Auerbach, and D.~P.~Arovas,
Phys. Rev. B {\bf 88}, 235108 (2013).

% NPRG calculation of Higgs in 2+1 dimensions
\bibitem{Ran14}
A.~Ran\c{c}on and N.~Dupuis, 
Phys. Rev. B {\bf 89}, 180501(R) (2014).

% more detailed NPRG calculation, figure including 1/w behavior
\bibitem{Ros15}
F.~Rose, F.~L\'{e}onard, and N.~Dupuis, 
Phys. Rev. B {\bf 91}, 224501 (2015).

% Higgs spectral function in 4-epsilon dimensions
\bibitem{Kat15}
Y.~T.~Katan and D.~Podolsky,
Phys. Rev. B {\bf 91}, 075132 (2015).

% Higgs in cold atomic gases
\bibitem{End12}
M.~Endres, T.~Fukuhara, D.~Pekker, M.~Cheneau, P.~Schau\ss, Ch.~Gross, 
E.~Demler, S.~Kuhr, and I.~Bloch, Nature {\bf 487}, 454 (2012).

%crystal growth
\bibitem{Nak01}
S.~Nakatsuji and Y.~Maeno,
J. Solid State Chem. {\bf 156}, 26 (2001).

% undoped, phonons assigned and analyzed		
\bibitem{Rho05}
H.~Rho, S.~L.~Cooper, S.~Nakatsuji, H.~Fukazawa, and Y.~Maeno,
Phys. Rev. B {\bf 71}, 245121 (2005).

% optics Ca214, gap 0.5 eV
\bibitem{Jun03}
J.~H.~Jung, Z.~Fang, J.~P.~He, Y.~Kaneko, Y.~Okimoto, and Y.~Tokura, 
Phys. Rev. Lett. {\bf 91}, 056403 (2003).

% pressure dependent phonons and 2M
\bibitem{Sno02}
C.~S.~Snow, S.~L.~Cooper, G.~Cao, J.~E.~Crow, H.~Fukazawa, 
S.~Nakatsuji, and Y.~Maeno, 
Phys. Rev. Lett. {\bf 89}, 226401 (2002).

% doping dependent phonons
\bibitem{Rho03}
H.~Rho, S.~L.~Cooper, S.~Nakatsuji, H.~Fukazawa, and Y.~Maeno,
Phys. Rev. B {\bf 68}, 100404(R) (2003).

% approach utilizing Green's functions - cited in Klein's chapter
\bibitem{Nit74}
A.~Nitzan, Mol.~Phys. {\bf 27}, 65 (1974).

% Fano background extraction
\bibitem{Kle75}
M.~V.~Klein, in {\it Light Scattering in Solids}, edited by M.~Cardona
(Springer-Verlag, Heidelberg, 1975).

\bibitem{Che93}
X.~K.~Chen, E.~Altendorf, J.~C.~Irwin, R.~Liang, and W.~N.~Hardy,
Phys. Rev. B {\bf 48}, 10530 (1993).	

\bibitem{noteBetts}
Clusters 16A, 18A, and 20A from D.~D.~Betts, H.~Q.~Lin, and J.~S.~Flynn, 
Can. J. Phys. {\bf 77}, 353 (1999) were used. The spectra were broadened 
by gaussians ($\mathrm{FWHM}$=$8\:\mathrm{meV}$), apart from the 
low-energy $B_{1g}$ peak whose lineshape is taken from the $B$ 
feature in Fig.~\ref{fig:fit}(a).

\bibitem{Fle68}
P.~A.~Fleury and R.~Loudon, Phys. Rev. {\bf 166}, 514 (1968).

\bibitem{notezz}
Note that a direct excitation of the magnetic continuum requires
$\vc{\epsilon}_\mathrm{in}$ and $\vc{\epsilon}_\mathrm{out}$ to be in the
$xy$-plane. We have verified that the feature $A$ and the associated Fano
asymmetries are indeed absent in the $zz$-polarized Raman spectra (not shown).

% two-Higgs above two-magnon peak in B1g Raman:
\bibitem{Wei15}
S.~A.~Weidinger and W.~Zwerger, 
Eur. Phys. J. B {\bf 88}, 237 (2015).

% another INS
\bibitem{Kun15}
S.~Kunkem\"oller, D.~Khomskii, P.~Steffens, A.~Piovano, A.~A.~Nugroho,
and M.~Braden, Phys. Rev. Lett. {\bf 115}, 247201 (2015).

\end{thebibliography}
\end{document}